\documentclass{aa}  

\usepackage{graphicx}
\usepackage{txfonts}
\usepackage{hyperref}
\hypersetup{colorlinks=true,allcolors=[rgb]{0,0,0.5}}

\makeatletter
\renewcommand*\aa@pageof{, page \thepage{} of \pageref*{LastPage}}
\makeatother

\newcommand*\Eval[3]{\left.#1\right\rvert_{#2}^{#3}}

\begin{document} 

   \title{The effect of thermal non-equilibrium on kinetic nucleation}


   \author{S. Kiefer \inst{1, 2, 3, 4}
          \and
          D. Gobrecht \inst{1,5}
          \and
          L. Decin \inst{1}
          \and
          Ch. Helling \inst{3, 4}}

   \institute{Institute of Astronomy, KU Leuven, Celestijnenlaan 200D, 3001 Leuven, Belgium\\
              \email{sven.kiefer@kuleuven.be}
         \and
             Centre for Exoplanet Science, University of St Andrews, North Haugh, St Andrews, KY169SS, UK
         \and
             Space Research Institute, Austrian Academy of Sciences, Schmiedlstrasse 6, A-8042 Graz, Austria
         \and
             TU Graz, Fakult\"at f\"ur Mathematik, Physik und Geod\"asie, Petersgasse 16, A-8010 Graz, Austria
        \and Department of Chemistry \& Molecular Biology, University of Gothenburg, 40530
             G\"oteborg, Sweden     
             }

   \date{Received ...; accepted ...}

  \abstract
   {Nucleation is considered to be the first step in dust and cloud formation in the atmospheres of asymptotic giant branch (AGB) stars, exoplanets, and brown dwarfs. In these environments dust and cloud particles grow to macroscopic sizes when gas phase species condense onto cloud condensation nuclei (CCNs). Understanding the formation processes of CCNs and dust in AGB stars is important because the species that formed in their outflows enrich the interstellar medium. Although widely used, the validity of chemical and thermal equilibrium conditions is debatable in some of these highly dynamical astrophysical environments.
   }
   {We aim to derive a kinetic nucleation model that includes the effects of thermal non-equilibrium by adopting different temperatures for nucleating species, and to quantify the impact of thermal non-equilibrium on kinetic nucleation}
   {Forward and backward rate coefficients are derived as part of a collisional kinetic nucleation theory ansatz. The endothermic backward rates are derived from the law of mass action in thermal non-equilibrium. We consider elastic collisions as thermal equilibrium drivers.}
   {For homogeneous TiO$_2$ nucleation and a gas temperature of 1250 K, we find that differences in the kinetic cluster temperatures as small as 20 K increase the formation of larger TiO$_2$ clusters by over an order of magnitude. Conversely, an increase in cluster temperature of around 20 K at gas temperatures of 1000 K can reduce the formation of a larger TiO$_2$ cluster by over an order of magnitude. 
   }
   {
   Our results confirm and quantify the prediction of previous thermal non-equilibrium studies. Small thermal non-equilibria can cause a significant change in the synthesis of larger clusters. Therefore, it is important to use kinetic nucleation models that include thermal non-equilibrium to describe the formation of clusters in environments where even small thermal non-equilibria can be present.
   }

   \keywords{Astrochemistry -- Methods: analytical -- Planets and satellites: atmospheres -- Stars: AGB and post-AGB
             }

   \maketitle
%

\section{Introduction}
\label{sec:Introduction}
    Signatures of active dust formation are present in many astrophysical environments. For example, asymptotic giant branch (AGB) and Wolf–Rayet (WR) stars have been studied as dust-producing environments \citep{williams_infrared_1987, winters_circumstellar_1994, winters_circumstellar_1995, woitke_dust_2000, ferrarotti_composition_2006, hofner_dust_2009, gail_formation_2010, karovicova_new_2013, gobrecht_dust_2016, gupta_thermodynamics_2020}. Dust produced within these environments is ejected into the interstellar medium (ISM) by radiation pressure on dust particles and thus plays an important role in the chemical enrichment of the ISM \citep{matsuura_global_2009, ventura_gas_2020}. In exoplanets, observations of mass-losing atmospheres \citep[e.g.][]{lieshout_dusty_2014} and exoplanet atmospheres \citep[e.g.][]{kreidberg_clouds_2014} have shown evidence of dust and clouds, respectively. Various models have been developed to describe the formation of cloud particles or their effects on exoplanet and brown dwarf atmospheres \citep[e.g.][]{tsuji_evolution_1996, tsuji_dust_2002, tsuji_dust_2005, allard_limiting_2001, allard_model_2003, ackerman_precipitating_2001, woitke_dust_2003, helling_dust_2008}. Because of the large cloud opacities in the optical wavelength regime, transmission spectra obtained from cloudy planets are typically featureless at those wavelengths \citep{pont_detection_2008, pont_prevalence_2013, bean_ground-based_2010, crossfield_warm_2013, knutson_hubble_2014, sing_hst_2015, sing_continuum_2016}. 
   
    Nucleation describes the clustering of gas-phase species to cloud condensation nuclei (CCNs). One way of describing CCN formation is classical nucleation theory (CNT). This theory assumes chemical and thermal equilibrium of the gas phase, and that the Gibbs free energy of formation of the nucleating species can be approximated by macroscopic properties \citep[see e.g.][]{helling_modelling_2013}. Modified classical nucleation theory (MCNT) extends CNT by connecting the surface tension with the Gibbs free energy of the nucleating cluster species \citep{draine_timedependent_1977, gail_dust_1984, lee_dust_2018}. 
    
    Recently, \citet{tielens_dust_2022} pointed out the importance of kinetics for dust formation in astrophysical environments. In order to study nucleation in chemical non-equilibrium, kinetic reaction networks including nucleation theory can be used \citep[e.g.][]{girshick_kinetic_1990, kalikmanov_self-consistent_1993, patzer_dust_1998, boulangier_developing_2019,gobrecht_bottom-up_2022}. Starting from the smallest entities of a substance corresponding to one stoichiometric formula unit (a monomer; e.g. TiO$_2$), these particles react to form larger (sub-)nanometre-sized structures (e.g. TiO$_2$ + (TiO$_2$)$_8$ $\rightarrow$ (TiO$_2$)$_9$). These larger particles are subsequently referred to as clusters or N-mers. For small clusters the nucleation is likely to proceed as termolecular reactions involving a third body. Conversely, larger clusters can dissociate into smaller clusters or monomers at sufficiently high temperatures (e.g. (TiO$_2$)$_9$ $\rightarrow$ TiO$_2$ + (TiO$_2$)$_8$). In many cases, the dissociations are induced by collisions.
    
    The Becker-D\"oring equations \citep{becker_kinetische_1935, burton_nucleation_1977, penrose_chapter_1979} describe kinetic nucleation if cluster growth and dissociation happens solely via monomers and the same chemical species. Their framework was later extended to polymer nucleation \citep{ball_discrete_1990, carr_asymptotic_1992, carr_asymptotic_1994}. We note that the nucleation can also proceed via cluster stoichiometries that are different from the crystalline bulk (i.e. the mineral).
    
    Constructing comprehensive chemical kinetic networks for modelling dust nucleation and growth in stellar outflows is a complex task \citep{gail_inorganic_1998, gail_mineral_1999, plane_nucleation_2013, gobrecht_dust_2016, bromley_under_2016, gobrecht_bottom-up_2022}. In addition to the challenges in modelling chemical non-equilibrium, the assumption of thermal equilibrium has been criticised \citep{donn_does_1985, goeres_chemistry_1996, sedlmayr_formation_1997, ferrarotti_mineral_2002}. Recent observations of AGB stars \citep{fonfria_detailed_2008, fonfria_abundance_2017, fonfria_multifrequency_2021} confirmed the presence of thermal non-equilibrium.
    
    \citet{plane_master_2022} analysed the effect of vibrational non-equilibria on the dissociation rate of silicate clusters, OSi(OH)$_2$, in stellar outflows. They found that the corresponding dissociation rate of OSi(OH)$_2$ is reduced by several orders of magnitude in vibrational non-equilibrium. Reactions of Ca, Fe, and Mg with OSi(OH)$_2$ might represent promising pathways to create metal silicon oxide clusters (e.g. CaSiO$_3$, FeSiO$_3$, and MgSiO$_3$). Combining the thermal non-equilibrium dissociation rate of OSi(OH)$_2$ with the chemical network of \citet{plane_nucleation_2013} showed that the reduced dissociation rate of OSi(OH)$_2$ increased the abundance of Ca-, Fe-, and Mg-bearing silicate clusters. This chemical network also includes TiO$_2$ which can form Ca, Fe, and Mg titanates via OTi(OH)$_2$. The impact of thermal non-equilibrium on Ti- and Si-bearing nucleating species is considered to be important for dust formation \citep[e.g.][]{waters_mineralogy_1996, gail_mineral_1999, goumans_stardust_2013, lee_dust_2015, gobrecht_dust_2016, bromley_under_2016, boulangier_developing_2019, sindel_revisiting_2022}.
    
    Thermal non-equilibrium is not only important for small molecular dust precursors but also for larger clusters \citep{nuth_effects_1985}. \citet{nuth_joseph_a_iii_silicates_2006} analysed the impact of vibrational non-equilibrium on the nucleation of SiO. They used the vibrational temperature of a single SiO molecule to approximate the vibrational temperatures of larger (SiO)$_N$ clusters \citep{nuth_vibrational_1981}. Using this approximation, they analysed the applicability of CNT in circumstellar environments and found that the presence of vibrational disequilibrium enables SiO dust formation at higher kinetic temperatures. Although the authors conclude that CNT cannot be made to work in expanding circumstellar shells, CNT is used owing to the lack of a suitable alternative. Even though the importance of thermal non-equilibrium on titania and silica dust precursors has been shown, only a few modelling attempts have been made to include thermal non-equilibrium in kinetic nucleation models \citep{patzer_dust_1998, lazzati_non-local_2008, kohn_dust_2021, plane_master_2022}. These models start from thermal equilibrium conditions and include two different temperatures, one for the gas phase species and one for the clusters or dust.
    
    In this study, we test and quantify the predicted importance of thermal non-equilibrium on kinetic nucleation. We prescribe different temperatures for each considered cluster size and we study the effect of different kinetic (translational) temperatures. We aim to address internal (vibrational and rotational) thermal non-equilibrium in more detail in a future study.
    
    This paper is organized as follows. We derive a kinetic nucleation framework from first principles including thermal and chemical non-equilibrium (Section \ref{sec:Model}). In Section \ref{sec:RelaxTime}, relaxation timescales for collisional and radiative cooling and heating processes are analysed in order to determine the importance of thermal non-equilibrium. We then use our model and recently published thermodynamic data to quantify the effects of thermal non-equilibrium on nucleation in Section \ref{sec:Testing}. Lastly, we present our conclusions in Section \ref{sec:Conclusion}.

\section{Kinetic nucleation model}
\label{sec:Model}

    In this section we describe the kinetic nucleation model. In Section \ref{sec:Model_KinInter}, the approach of \citet{boulangier_developing_2019} is used to show how the chemical reaction network formalism can be used to describe polymer nucleation. In Section \ref{sec:Model_RelSpeedDist}, the derivation of the Maxwell-Boltzmann relative speed distribution in thermal non-equilibrium is made and in Section \ref{sec:Model_forward_reaction_rate}, \ref{sec:Model_backward}, and \ref{sec:Model_backward} the forward and backward reaction rates are derived.

    \subsection{Kinetic reactions}
    \label{sec:Model_KinInter}
    
    Each cluster consists of $N$ basic building blocks (e.g. TiO$_2$) that are linked by chemical bonds (e.g. (TiO$_2$)$_N$). The change in cluster number densities $n_N$ [cm$^{-3}$] can be described by the following coupled ordinary differential equations (ODEs) \citep[Eq. 1 of ][]{boulangier_developing_2019}:
    \begin{align}
        \label{eq:chemical_network}
        \frac{d n_N}{d t} = \sum_{j \in F_N} \left( k_j^+ \prod_{r \in R^+_j} n_r \right) - \sum_{j \in \mathcal{D}_N} \left( k_j^- \prod_{r \in R^-_j} n_r \right) ,
    \end{align}
    where $F_N$ is the set of forward reactions ($R_j^+$) and $\mathcal{D}_N$ is the set of backward reactions ($R_j^-$). In the following, $R_j$ is defined as either $R_j^+$ or $R_j^-$ and each variable specific to reaction $R_j$ has a $j$ subscript. Similarly, $k_j$ is either $k_j^+$ or $k_j^-$. The variable $n_r$ [cm$^{-3}$] denotes the number densities of the cluster involved in reaction $R_j$ and $k_j$ [cm$^{3(J-1)}$s$^{-1}$] is the reaction rate for the reaction $R_j,$ where $J$ is the number of reactants in reaction $R_j$. For small cluster sizes and low densities, three-body association reactions are the dominant cluster nucleation process (see e.g. \citep{bromley_under_2016}). The reaction rates for termolecular associations and their reverse collisional dissociation are discussed in Section \ref{sec:Model_3body}. For larger clusters, two-body association reactions ($J$=2) are assumed to be the dominant forward reaction for which the reaction rate $k_j^+$ is the following \citep[based on ][]{peters_chapter_2017}:
    \begin{align}
        \label{eq:k}
        k_j^+ = \int_0^{\infty} \alpha_j(\nu_r) ~  \sigma_j(\nu_r) ~ \nu_r ~ f(\nu_r) ~ d \nu_r,
    \end{align}
    where $\alpha_j(\nu_r)$ is the sticking coefficient, $\sigma_j(\nu_r)$ [cm$^2$] the reaction cross section, $\nu_r$ [cm s$^{-1}$] the relative velocity of the colliding particles, and $f (\nu_r)$ the Maxwell-Boltzmann velocity distribution (see Sect. \ref{sec:Model_RelSpeedDist}). Due to the lack of data on sticking coefficients for nucleating species, we set $\alpha_j(\nu_r) = 1$ for the rest of this paper. Therefore, all reaction rates are upper limits. This is a frequently used approximation for the sticking coefficient \citep[e.g.][]{lazzati_non-local_2008, bromley_under_2016, boulangier_developing_2019}.
    
    The cross section is a measure of the probability that a cluster reacts with other clusters\footnote{In the case of kinetic chemistry, the cross section is a statistical property. Therefore, for non-spherical particles, the cross section $\sigma$ has to be found by averaging the angle-dependent cross section $\sigma(\phi, \theta)$ over the azimuthal angles $\phi$ and polar angles $\theta$.}. In the most general case, it can depend on the relative velocity $\nu_r$. For small clusters, long-range interactions from electrostatic forces need to be taken into account\footnote{A comparison of van der Waals radii to geometric radii for TiO$_2$ can be found in Appendix \ref{sec:Appendix_Gibbs_TiO2_data} or in \citet{kohn_dust_2021}.} \citep{bromley_computational_2016, kohn_dust_2021}. We describe the cross section by a collision of two hard spheres:
    \begin{align}
        \label{eq:def_inelasitcCorssSec}
        \sigma_j = \pi (r_1 +r_2)^2,
    \end{align}
    where $r_1$ and $r_2$ [cm] are the interaction radii of the collision partners. We provide interaction radii for (TiO$_2$)$_N$ clusters (Table \ref{tab:tio2_data}), which are used in our reference nucleation case.
    
    The goal of the present study is to assess the impact of thermal non-equilibrium on the kinetic formation of clusters. The gas is assumed to be in thermal equilibrium, but the clusters might not be. Clusters of a given size $N$ are described by a kinetic cluster temperature ($T^\mathrm{kin}_N$) and an internal cluster temperature ($T^\mathrm{int}_N$) which can differ from each other and from the gas phase temperature $T_\mathrm{gas}$. The internal temperature includes vibrational and rotational contributions which are not differentiated further in the present study.
    
    An $N$-mer is composed of $N x_1$ atoms, where $x_1$ is the number of atoms in a monomer unit. Therefore an $N$-mer has $3 N x_1$ degrees of freedom of which three describe translational movement and $D^f_N = 3 N x_1 - 3$ describe internal degrees of freedom (including three rotations and $3 N x_1 - 6$ vibrations). Using the equipartition theorem, we can define
    \begin{align}
        \label{eq:Thermodynmaic_Temperature}
        E_N^\mathrm{tot} = E_N^\mathrm{kin} + E_N^\mathrm{int} = \frac{3}{2} k T_N^\mathrm{kin} + \frac{D_N^f}{2} k T_N^\mathrm{int},
    \end{align}
    where $k = 1.381 \times 10^{-16}$ erg K$^{-1}$ is the Boltzmann constant. In our model, each cluster size can be at a different kinetic temperature $T_N^\mathrm{kin}$ and at a different internal temperature $T_N^\mathrm{int}$.

    \subsection{Two-particle speed distribution}
    \label{sec:Model_RelSpeedDist}
    
    To derive the relative speed distribution for two particle ensembles ($P_1$ and $P_2$) with different masses ($m_1$ and $m_2$) and temperatures ($T_1$ and $T_2$), we follow the derivation of \citet{kusakabe_relative_2019}. We use an adapted version of their centre of mass transformation, which we call the temperature-weighted centre of mass (TCM). We perform the TCM transformation as follows:
    \begin{align}
        M_T &\equiv \frac{m_1}{T_1^\mathrm{kin}} + \frac{m_2}{T_2^\mathrm{kin}} = \frac{m_1 T_2^\mathrm{kin} + m_2 T_1^\mathrm{kin}}{T_1^\mathrm{kin} T_2^\mathrm{kin}} ,
    \end{align}
    \begin{align}
        \mu &\equiv \frac{m_1 m_2}{m_1 + m_2},
    \end{align}
    \begin{align}
        \mu_T &\equiv \frac{\frac{m_1}{T_1^\mathrm{kin}} \frac{m_2}{T_2^\mathrm{kin}}}{\frac{m_1}{T_1^\mathrm{kin}} + \frac{m_2}{T_2^\mathrm{kin}}} = \frac{m_1 m_2}{m_1 T_2^\mathrm{kin} + m_2 T_1^\mathrm{kin}} ,
    \end{align}
    \begin{align}
         \label{eq:Vr}
         \vec{\nu}_\mathrm{r} &\equiv  \vec{\nu}_1 -  \vec{\nu}_2 ,
    \end{align}
    \begin{align}
         \vec{v}_T &\equiv \frac{\frac{m_1}{T_1^\mathrm{kin}} \vec{\nu}_1 + \frac{m_2}{T_2^\mathrm{kin}} \vec{\nu_2}}{\frac{m_1}{T_1^\mathrm{kin}} +  \frac{m_2}{T_2^\mathrm{kin}}} = \frac{m_1 T_2^\mathrm{kin} \vec{\nu}_1 + m_2 T_1^\mathrm{kin}  \vec{\nu}_2}{m_1 T_2^\mathrm{kin} + m_2 T_1^\mathrm{kin}} ,
    \end{align}
    where $m_1$, $m_2$ [g] are the masses of the particles within $P_1$ and $P_2$, respectively. Furthermore, $T_1^\mathrm{kin}$, $T_2^\mathrm{kin}$ [K] are their kinetic temperatures and \vec{\nu_1}, \vec{\nu_2} [cm s$^{-1}$] their velocity. In addition, $M_T$ [g T$^{-1}$] is the temperature-weighted total mass, $\mu$ [g] the reduced mass, $\mu_T$ [g T$^{-1}$] the temperature-weighted reduced mass, $\vec{\nu}_\mathrm{r}$ [cm s$^{-1}$] the relative velocity, and $\vec{v}_T$ [cm s$^{-1}$] the TCM velocity. Using these definitions, we can find the following relation:
    \begin{align}
        \frac{m_1}{T_1^\mathrm{kin}}  \vec{\nu}_1^2 + \frac{m_2}{T_2^\mathrm{kin}}  \vec{\nu}_2^2 = M_T  \vec{v}_T^2 + \mu_T  \vec{\nu}_r^2 .
    \end{align}
    From here, we assume that the velocity distribution of the clusters follows the Maxwell-Boltzmann velocity distribution for a particle of mass $m$ [g]:
    \begin{align}
        \label{eq:MaxwellBoltzmann}
        f(\vec{\nu}) d\vec{\nu} = \left( \frac{m}{2 \pi k T} \right)^{3/2} \exp \left( - \frac{m \vec{\nu}^2}{2 k T} \right) d\vec{\nu}. 
    \end{align} 
    The relative velocity distribution $f_r (v_r)$ can be found using an equivalent derivation\footnote{The relative speed distribution can be found by convolution of the velocity distributions of the collision partners: $f_r ( \vec{\nu}_r) = \int \vec{d} \vec{\nu}_1 \int \vec{d} \vec{\nu}_2 f( \vec{\nu}_1) f(\vec{\nu}_2)  ~ \delta (|\vec{\nu_r}| - |\vec{\nu}_1 - \vec{\nu}_2|)$, where $\delta$ is the Dirac delta function.} as in Eq. 9 and 10 of \citet{kusakabe_relative_2019}:
    \begin{align}
        \int_{\mathbb{R}^3} f_r ( \vec{\nu}_r) d  \vec{\nu}_r &= \int_{\mathbb{R}^3} f( \vec{\nu}_1) d \vec{\nu}_1 \int_{\mathbb{R}^3} f(\vec{\nu}_2) d \vec{\nu}_2 \\
        \nonumber &= \left( \frac{1}{2 \pi k} \right)^3 \left( \frac{m_1 m_2}{T_1^\mathrm{kin} T_2^\mathrm{kin}} \right)^{3/2} \\ 
        & \quad \int_{\mathbb{R}^3} \int_{\mathbb{R}^3} \exp \left( -\frac{M_T  \vec{v}_T^2 + \mu_T  \vec{\nu}_r^2}{2 k } \right) d \vec{\nu}_r d \vec{v}_T \\
        &= \int_{\mathbb{R}^3} \left( \frac{\mu_T}{2 \pi k} \right)^{3/2} \exp \left( - \frac{\mu_T \vec{\nu}_r^2}{2 k}\right) d \vec{\nu}_r .
    \end{align}
    Assuming spherical symmetry ($d \vec{\nu}_r = 4 \pi \nu_r^2 d \nu_r $), this leads to the following relative speed distribution:
    \begin{align}
       \label{eq:MB_relSpeedDist}
       f_r (\nu_r) d\nu_r = \left(\frac{\mu_T}{2 \pi k} \right)^{3/2} 4 \pi \nu_r^2 \exp \left( - \frac{\mu_T \nu_r^2}{2 k}\right) d\nu_r,
    \end{align}
    which is similar to a Maxwell-Boltzmann distribution of the centre of mass frame, but the temperature-weighted reduced mass $\mu_T$ replaces the reduced mass. This distribution can be used to describe the relative velocity of two particle ensembles (e.g. a cluster of different sizes) at different temperatures. 
    
    For large temperature differences (e.g. $T_1^\mathrm{kin} \gg T_2^\mathrm{kin}$) or large mass differences (e.g. $m_1 \ll m_2$), $\mu_T$ becomes 
   \begin{align}
       \mu_T &\approx \frac{m_1}{T_1^\mathrm{kin}} ,
   \end{align}
   which recovers the Maxwell-Boltzmann distribution for the particle ensemble $P_1$. If the collision partners have the same kinetic temperature ($T_1^\mathrm{kin} = T_2^\mathrm{kin}$), we find   \begin{align}
       \mu_T &= \frac{\mu}{T} ,
   \end{align}
   which recovers the result for thermal equilibrium, defined as $T_\mathrm{gas} = T^\mathrm{kin}_N = T^\mathrm{int}_N$ \citep{boulangier_developing_2019}.

   \subsection{Forward reaction rate}
   \label{sec:Model_forward_reaction_rate}
   
   Combining Eq. \ref{eq:k}, \ref{eq:def_inelasitcCorssSec}, and \ref{eq:MB_relSpeedDist}, we find a general expression for the forward reaction rate $k^+_j$ of a two-body collision between particles from $P_1$ (with $m_1$ and $T_1$) and particles from $P_2$ (with $m_1$ and $T_1$) for a sticking coefficient of $\alpha = 1$:
   \begin{align}
        \label{eq:k_integrated}
        k^+_j&= \int_0^{\infty}  \pi (r_1 + r_2)^2 \nu_r \left(\frac{\mu_T}{2 \pi k} \right)^{3/2} 4 \pi \nu_r^2 \exp \left( - \frac{\mu_T \nu_r^2}{2 k}\right) d \nu_r \\
        \label{eq:MB_difficult}
        &= \pi (r_1 + r_2)^2 \sqrt{\frac{8 k}{\pi \mu_T}} = \pi (r_1 + r_2)^2 \langle \nu_r \rangle ,
   \end{align}
   where $\langle \nu_r \rangle = \sqrt{{8 k} / {\pi \mu_T}}$ is the average relative speed.

    \subsection{Association and dissociation of small clusters}
    \label{sec:Model_3body}
    
    For small cluster sizes (e.g. N$\le$4 for TiO$_2$) and gas densities n$_{gas}$=10$^{12}$-10$^{14}$ cm$^{-3}$ considered in this study, termolecular associations (A + B + M $\rightarrow$ AB + M) are the dominant cluster growth reactions, and they are favoured over bimolecular radiative associations (A + B $\rightarrow$ AB + h$\nu$) (see e.g. \citet{bromley_under_2016}). For the reverse collisional dissociation reactions (AB + M $\rightarrow$ A + B + M), we use accurate CCSD(T)/6-311+G(2d,2p) single point energies to calculate the dissociation energy $E_\mathrm{AB}$ [erg mol$^{-1}$]. The dissociation rate is calculated as
    \begin{align}
        k_\mathrm{3B}^- = A \exp \left( \frac{-E_\mathrm{AB}}{RT}\right) .
    \end{align}
    The dissociation rates are adjusted for thermal non-equilibrium conditions by accounting for the different relative velocities of the gas particles M and the clusters:
    \begin{align}
        k_\mathrm{3B}^- = A \exp \left( \frac{-E_\mathrm{AB}}{RT}\right) \sqrt{\frac{m_\mathrm{gas} T^\mathrm{kin}_\mathrm{AB} + m_\mathrm{AB} T_\mathrm{gas}} {(m_\mathrm{gas} + m_\mathrm{AB}) T_\mathrm{gas}}} .
    \end{align}
    We note that, in addition to the relative velocities, the clusters could have significantly cooler internal (vibrational) temperatures than the gas particles, which is not taken into account in the present study.
    
    The corresponding three-body forward association reaction rates for TiO$_2$ are calculated using detailed balance as described in Section \ref{sec:Model_backward} with cluster partition functions derived in \cite{sindel_revisiting_2022}. For the pre-exponential rate constant $A$ [cm$^{3}$s$^{-1}$], a value of $1.4 \times 10^{-9}$ cm$^{3}$s$^{-1}$ is used. This approximated value is based on an upper limit of $2 \times 10^{-9}$ cm$^{3}$s$^{-1}$ that is still considered to be physical (see e.g. \citet{gobrecht_bottom-up_2022}), and it is similar to the pre-exponential factors for collisional dissociations of aluminium oxide clusters \citep{catoire_kinetic_2003}. The dissociation rate coefficients for TiO$_2$ are shown in Table \ref{table:3_body_tio2} and example values for the association reaction rate coefficients are shown in Table \ref{table:3_body_association_tio2}. The detailed balance and equilibrium constant computations are approximated in kinetic-to-internal thermal equilibrium ($T_N^\mathrm{int} = T_N^\mathrm{kin} \neq T_\mathrm{gas}$).

   \subsection{Backward reaction rate}
   \label{sec:Model_backward}
    Backward reactions are considered as spontaneous processes that cause larger clusters to fragment into smaller clusters. For kinetic nucleation, a general relation can be written as $A + B \leftrightarrows C$. To determine the rate coefficients $k^-_j$ of this process, we make use of the principle of detailed balance \citep[Milne relation; Chapter 12.2.2 of ][]{gail_physics_2013} which assumes that, in detailed balance (e.g. chemical equilibrium), the ratio between the forward and backward rate of each reaction is the same. Therefore, in chemical equilibrium (marked with $\mathring{ }$ ), the species flux of each forward reaction is equivalent to the flux of the corresponding backward reaction:
    \begin{align}
        \label{eq:Back_keq}
        {k^+_j} \mathring{n}_A \mathring{n}_B  &= {k^-_j} \mathring{n}_{A+B},
    \end{align}
    where $\mathring{n}_A$, $\mathring{n}_B$, and $\mathring{n}_{A+B}$ [cm$^{-3}$] are the number densities in chemical equilibrium. Eq. \ref{eq:Back_keq} can be rewritten as
    \begin{align}
        {k^-_j} &= {k^+_j} \frac{\mathring{n}_A \mathring{n}_B}{\mathring{n}_{A+B}} .
    \end{align}

    To find the number densities in chemical equilibrium, the total Gibbs free energy (including translational, rotational, and vibrational contributions) in thermal non-equilibrium $G^{non-eq}$ [erg] is minimized. The setup considered here can be described by nucleating clusters immersed in an ambient gas. We assume a single non-clustering gas species and a single clustering species to simplify notation, but the derivation also holds for multiple non-clustering and clustering species.  Furthermore, we assume that all clusters other than the monomer can be in thermal non-equilibrium (in the special case of thermal equilibrium, both temperatures are equal to the gas temperature: $T_\mathrm{gas} = T_i^\mathrm{kin} = T_i^\mathrm{int}$) and each $i$-mer has a kinetic temperature $T_i^\mathrm{kin}$ and an internal temperature of $T_i^\mathrm{int}$.
    
    The Gibbs free energy of cluster\footnote{We assume that each metastable cluster structure quickly relaxes to their global minimum after formation. Therefore, we can use the Gibbs free energies of the global minimum. For a detailed analysis of relaxation timescales of the potential energy surface, the reader is referred to \citet{doye_potential_1996}.} where the internal cluster temperature $T_i^\mathrm{int}$ differs from the kinetic temperature $T_i^\mathrm{kin}$ can be written as follows:
    \begin{align}
        \label{eq:gibbssplit_1}
        G_i^{non-eq}(T_i^\mathrm{int}, T_i^\mathrm{kin}, p_i, N_i) = G_i(T_i^\mathrm{kin}, p_i, N_i) + N_i ~\omega_i (T_i^\mathrm{kin}, T_i^\mathrm{int}) ,
    \end{align}
    where $G^{non-eq}_i (T_i^\mathrm{kin}, T_i^\mathrm{int}, p_i, N_i)$ is the $i$-mer's Gibbs free energy in kinetic-to-internal thermal non-equilibrium (defined as $T^\mathrm{kin}_i$ $\neq$ $T^\mathrm{int}_i$), $p_i$ [dyn cm$^{-2}$] is the partial pressure of the $i$-mer and $N_i$ the number of $i$-mers. Furthermore, $\omega_i (T_i^\mathrm{kin}, T_i^\mathrm{int})$ [erg] is the difference in Gibbs free energy of a single $i$-mer in thermal equilibrium to the Gibbs free energy of a cluster in kinetic-to-internal thermal non-equilibrium. Because changes in the internal temperature do not affect the $i$-mer's partial pressure $p_i$, volume, or number densities of the gas, $\omega_i$ only depends on kinetic and internal temperature. Summing Eq. \ref{eq:gibbssplit_1} for all clusters and gas species, the total Gibbs free energy can be written as follows:
    \begin{align}
        \nonumber G^{non-eq}(T_0^\mathrm{int}, &... , T_r^\mathrm{int}, T_0^\mathrm{kin}, ... , T_r^\mathrm{kin}, p_0, ..., p_r, N_0, ..., N_r) \\
        \label{eq:splitgibs}
        &= \sum_{i=0}^r G_i^{non-eq}(T_i^\mathrm{int}, T_i^\mathrm{kin}, p_i, N_i) \\
        \label{eq:gibbsderi_g}
        &= \sum_{i=0}^r G_i(T_i^\mathrm{kin}, p_i, N_i) + N_i ~\omega_i (T_i^\mathrm{kin}, T_i^\mathrm{int}),
    \end{align}
    where the subscript 0 describes the gas ($T_\mathrm{gas} = T_0^\mathrm{int} = T_0^\mathrm{kin}$ and $N_\mathrm{gas} = N_0$), the subscript $i \ge 1$ represents the cluster sizes, and $r$ denotes the largest cluster considered. Next, we need to convert the partial pressures $p_i$ to the total pressure $p = \sum_{i=0}^r p_i$ [dyn cm$^{-2}$]: 
    \begin{align}
        G_i(T_i^\mathrm{kin}, p_i, N_i) &= G_i(T_i^\mathrm{kin}, p, N_i) + \int_p^{p_i} \frac{\partial G_i(T_i^\mathrm{kin},p',N_i)}{\partial p'} dp'\\
        &= G_i(T_i^\mathrm{kin}, p, N_i) + \int_p^{p_i} \frac{N_i k T_i^\mathrm{kin}}{p'} dp' \\
        \label{eq:gibbsderi_gi}
        &= G_i(T_i^\mathrm{kin}, p, N_i) + N_i k T_i^\mathrm{kin} \ln \left( \frac{N_i}{N}\right),
    \end{align}
    where $N = \sum_{i=0}^r N_i$ is the total number of particles (gas and clusters). Combining Eq. \ref{eq:gibbsderi_g} and \ref{eq:gibbsderi_gi} as well as the definition of the chemical potential $\mu$ ($G = N \mu$) leads to the Gibbs free energy of the mixture of gas and clusters:
    \begin{align}
        \label{eq:full_non-eq_gibbs}
        \nonumber &G^{non-eq}(T_0^\mathrm{int}, ... , T_r^\mathrm{int}, T_0^\mathrm{kin}, ... , T_r^\mathrm{kin}, p, N_0, ..., N_r) \\
        \nonumber&= N_\mathrm{gas} \mu_\mathrm{gas}(T_\mathrm{gas}, p) + N_\mathrm{gas} k T_\mathrm{gas} \ln \left( \frac{N_\mathrm{gas}}{N}\right) +  \sum_{i=1}^r N_i \mu_i (T_i^\mathrm{kin}, p)  \\
        &\quad \quad \quad + N_i k T_i^\mathrm{kin} \ln \left( \frac{N_i}{N}\right) + N_i ~\omega_i (T_i^\mathrm{kin}, T_i^\mathrm{int}),
    \end{align}
    where $\mu_i (T_i^\mathrm{kin}, p)$ [erg] is the chemical potential of the clusters and $\mu_\mathrm{gas} (T_\mathrm{gas}, p)$ [erg] the chemical potential of the gas species\footnote{In the literature usually the molar Gibbs free energy $G_\mathrm{mol}$ is given, which is related to the chemical potential $\mu$ through the Avogadro constant $N_A,$ such that $N_A \mu = G_\mathrm{mol}$.}. This equation describes the Gibbs free energy of a mixture of gas and clusters in thermal non-equilibrium.
    
    To minimize the Gibbs free energy, the Lagrangian function can be used\footnote{The Lagrangian function is a tool for mathematical optimization problems in which equality constrains can be expressed using the Lagrangian multiplier $\lambda$.}. As an additional constraint, we assume that the number of basic building blocks (e.g. TiO$_2$) is conserved:
    \begin{align}
        \label{eq:def_lagrangian_constraint}
        C = \sum_{i=1}^r i~N_i. 
    \end{align}
    Using Eq. \ref{eq:full_non-eq_gibbs} and Eq. \ref{eq:def_lagrangian_constraint}, the Lagrangian function can be defined as
    \begin{align}
        \nonumber \mathcal{L} &= N_\mathrm{gas} \mu_\mathrm{gas}(T_\mathrm{gas}, p) + N_\mathrm{gas} k T_\mathrm{gas} \ln \left( \frac{N_\mathrm{gas}}{N}\right) - \lambda C \\
        & \nonumber \quad + \sum_{i=1}^r N_i \mu_i(T_i^\mathrm{kin}, p) + N_i k T_i^\mathrm{kin} \ln \left( \frac{N_i}{N}\right) \\
        & \quad \quad \quad + N_i ~\omega_i (T_i^\mathrm{kin}, T_i^\mathrm{int}) + \lambda i N_i .
    \end{align}
    The number of clusters N$_i$ are realistically much smaller than the total number of gas particles, such that 
    \begin{align}
        N_\mathrm{gas} &\gg \sum_{i=1}^r N_{i}.
    \end{align}
    Furthermore, we assume that the differences in temperature are small enough so as to not influence this approximation:
    \begin{align}
        N_\mathrm{gas} T_\mathrm{gas} &\gg \sum_{i=1}^r N_i T_i^\mathrm{kin} .
    \end{align}
    Taking the partial derivatives of the Lagrangian with respect to $N_1$, $N_j$ (for $j \geq 2$) and $\lambda$ and using the above approximations, we arrive at the following system of equations:
    \begin{align}
        \nonumber \frac{\partial \mathcal{L}}{\partial N_1} &= \mu_1 (T_\mathrm{gas}, p) + k T_\mathrm{gas} \ln \left( \frac{N_1}{N} \right) \\
        & \quad + \lambda + k T_\mathrm{gas} \frac{N - N_1}{N} - \sum_{\substack{i=0 \\ i \neq 1}}^r \frac{N_i k T_\mathrm{i}^\mathrm{kin}}{N} \\
        \label{eq:Lagrangian_dngas}
        &\approx \mu_\mathrm{1} (T_\mathrm{gas}, p) + k T_\mathrm{gas} \ln \left( \frac{N_1}{N} \right) + \lambda , \\
        \label{eq:Lagrangian_dni}
        \nonumber \frac{\partial \mathcal{L}}{\partial N_j} &= \mu_j(T_j^\mathrm{kin}, p) + k T_j^\mathrm{kin} \ln \left( \frac{N_j}{N} \right) + \omega_j (T_j^\mathrm{kin}, T_j^\mathrm{int}) \\ 
        & \quad + \lambda j x_{1} + k T_j^\mathrm{kin} \frac{N - N_j}{N} - \sum_{\substack{i=0 \\ i \neq j}}^r \frac{N_i k T_i^\mathrm{kin}}{N} , \\
        \label{eq:Lagrangian_dni_approx}
        &\nonumber \approx \mu_j(T_j^\mathrm{kin}, p) + k T_j^\mathrm{kin} \ln \left( \frac{N_j}{N} \right) + \omega_j (T_j^\mathrm{kin}, T_j^\mathrm{int}) \\ 
        & \quad + \lambda j + k (T_j^\mathrm{kin} - T_\mathrm{gas}) , \\
    \end{align}
    where $j = \{2,3,4, ... , r\}$. Minimizing the Lagrangian function with regard to $N_j$ is equivalent to setting  Eq. \ref{eq:Lagrangian_dni_approx} to 0 and this represents the state of chemical equilibrium. Setting Eq. \ref{eq:Lagrangian_dni_approx} to zero leads to
    \begin{align}
        \label{eq:Lagrangian_nn}
        \nonumber \frac{\mathring{N}_j}{\mathring{N}_\mathrm{gas}} = &\exp \left( \frac{-\mu_j(T_j^\mathrm{kin}, p)}{k T_j^\mathrm{kin}} \right) \\
        &\exp\left( \frac{- k (T_j^\mathrm{kin} - T_\mathrm{gas}) - \omega_j (T_j^\mathrm{kin}, T_j^\mathrm{int}) - \lambda j}{k T_j^\mathrm{kin}} \right) .
    \end{align}
    Using Eq. \ref{eq:Lagrangian_nn} for clusters of size $A$, $B,$ and $A+B$ allows us to write the following:
    \begin{align}
        \label{eq:Lagrangian_3n}
        \nonumber \frac{\mathring{N}_A \mathring{N}_B}{\mathring{N}_{A+B}} &= N_\mathrm{gas} \exp \left( \sum_{i \in \zeta} \frac{\delta(i)}{k T_i^\mathrm{kin}} \mu_i(T_i^\mathrm{kin}, p) \right) \\
        & \exp \left( \sum_{i \in \zeta} \frac{\delta(i)}{k T_i^\mathrm{kin}} \left[  k (T_i^\mathrm{kin} - T_\mathrm{gas}) + \omega_i (T_i^\mathrm{kin}, T_i^\mathrm{int}) + \lambda i \right] \right) ,
    \end{align}
     where $\zeta = \{ A, B, (A$+$B) \}$ defines the set of all involved cluster sizes and $\delta(i)$ is equal to 1 for products (here A+B) and -1 for the reactants (here A and B). To find the Lagrangian multiplier, we use Eq. \ref{eq:Lagrangian_dngas} and find
    \begin{align}
        \label{eq:Lagrangian_multiplier}
        \lambda = - \mu_1(T_\mathrm{gas}, p) - k T_\mathrm{gas} \ln \left( \frac{N_1}{N} \right).
    \end{align}
    Furthermore, it is convenient to express Eq. \ref{eq:Lagrangian_multiplier} in terms of the chemical potential ($\mu^\ominus(T_\mathrm{gas}) = \mu_i(T_\mathrm{gas}, p^\ominus)$) at standard pressure ($p^\ominus = 10^6$ dyn $cm^{-2}$). Doing this and using Eq. \ref{eq:Back_keq}, \ref{eq:Lagrangian_3n}, and \ref{eq:Lagrangian_multiplier} leads to the following backward reaction rate: 
    \begin{align}
        \label{eq:Gibbs_eq_general}
        \nonumber k^- = \frac{k^+ p^\ominus}{k T_\mathrm{gas}} \exp \Bigg( &\sum_{i \in \zeta} \frac{\delta(i)}{k T_i^\mathrm{kin}} \Big[ \mu_i^\ominus(T_i^\mathrm{kin}) -  i \mu^{\ominus}_1(T_\mathrm{gas}) + k (T_i^\mathrm{kin} - T_\mathrm{gas}) \\
        &   + \omega_i (T_i^\mathrm{kin}, T_i^\mathrm{int}) \Big] \Bigg) \left( \frac{k T_\mathrm{gas} \mathring{n}_1}{p^\ominus}\right)^{- \sum_{i \in \zeta} \delta(i) i T_\mathrm{gas} / T_i^\mathrm{kin}} .
    \end{align}

\section{Cooling and heating processes}
    \label{sec:RelaxTime}
    Collisional and radiative heating and cooling processes impact the cluster temperature. We first look at the collisional relaxation timescale of kinetic temperature $T^\mathrm{kin}_N$ to the gas temperature $T_\mathrm{gas}$ in Section \ref{sec:RelaxTime_kinetic}. Afterwards, we investigate the timescale of changes in internal temperature $T^\mathrm{int}_N$ via collisions in section \ref{sec:RelaxTime_internal} and via radiative processes in section \ref{sec:RelaxTime_Radiative}.

    \subsection{Kinetic temperature}
    \label{sec:RelaxTime_kinetic}
    
    To derive the collision-induced change in cluster kinetic temperature $T_N^\mathrm{kin}$, we follow the derivation of \citet{gail_physics_2013}. We assume that the energy redistribution happens through elastic and isotropic collisions between gas particles and the clusters. Considering an elastic and isotropic collision of two particles $P_1$ (with mass $m_1$) and $P_2$ (with mass $m_2$), the average ratio between the energy before and after the collision is the following \citep[Eq. 6.6 of ][]{gail_physics_2013}:
    \begin{align}
        \frac{E_1^\mathrm{after} - E_2^\mathrm{after}}{E_1^\mathrm{before} - E_2^\mathrm{before}} = 1 - \frac{8}{3} \frac{m_1 m_2}{(m_1 + m_2)^2} .
    \end{align}
    Every collision decreases the energy difference by that amount. The relaxation timescale is defined as the number of collisions $K_\mathrm{col}$ needed to reduce the kinetic energy of particles 1 with respect to particles 2 to $1/e$ of its initial value. Therefore, $K_\mathrm{col}$ is \citep[Eq. 6.9 of ][]{gail_physics_2013}
    \begin{align}
        K_\mathrm{col} &= -\ln \left( 1 - \frac{8}{3}\frac{m_1 m_2}{(m_1 + m_2)^2} \right)^{-1} \approx \frac{3}{8}\frac{m_1}{m_2} ,
    \end{align}
    where the approximation is the first term of a Taylor expansion and holds in the case of $m_1 \gg m_2$. This approximation holds for H$_2$ ($m = 2.02 ~u$) gas, including TiO$_2$ clusters ($m = 79.87 ~u$). To find the kinetic cooling timescale $\tau^\mathrm{kin}_\mathrm{gc}$ of elastic collisions, we multiply the number of collisions with $\langle t_\mathrm{gc}\rangle$ being the average time for a cluster (e.g. TiO$_2$) to collide with a gas particle (e.g. H$_2$):
    \begin{align}
        \langle t_\mathrm{gc} \rangle &= \frac{1}{n_\mathrm{gas} \sigma_\mathrm{N} \langle v_r \rangle} = \frac{1}{n_\mathrm{gas} \pi r_\mathrm{N}^2} \sqrt{\frac{\pi \mu_T}{8 k}} , \\
        (\tau^\mathrm{kin}_\mathrm{gc})^{-1} &= \frac{1}{K_\mathrm{col} * \langle t_{gc} \rangle} \\
        \label{eq:cooling_kinetic_collisional}
        &\approx \frac{8 m_\mathrm{gas}}{3 m_{N}}  n_\mathrm{gas} \pi r_{N}^2 \sqrt{\frac{8 k T_\mathrm{gas}}{\pi m_\mathrm{gas}}} ,
    \end{align}
    where the approximation is justified for $m_\mathrm{gas} \ll m_\mathrm{N}$. Looking at the reference case of (TiO$_2$)$_2$ clusters in a gas as described in Table \ref{table:Standard_example}, we find a collisional cooling timescale for the (TiO$_2$)$_2$ kinetic temperature of $\tau^\mathrm{kin}_\mathrm{gc} = 0.018$ s. We note that $\tau^\mathrm{kin}_\mathrm{gc}$ is inversely proportional to the gas number density $n_\mathrm{gas}$. Therefore, clusters residing in low-density regions (where collisions are not efficient enough to maintain thermal equilibrium) might not be in kinetic-to-gas thermal equilibrium (defined as $T^\mathrm{kin}_N = T_\mathrm{gas}$).
    
    \begin{table}
        \caption{Summary of the quantities used for the (TiO$_2$)$_2$ cooling timescale examples.} 
        \label{table:Standard_example}      
        \centering 
                \begin{tabular}{l l | l l}
                        \hline\hline
                        gas species & H$_2$          & Nucleating species &  TiO$_2$    \\
                        \hline
                        $T_\mathrm{gas}$ [K] & 1000  & $T_\mathrm{TiO2}$ [K] & 1000     \\
                        $n_\mathrm{gas}$ [cm$^{-3}$] & 10$^{12}$ & $n_\mathrm{TiO2}$ [cm$^{-3}$]  & 10$^{4}$  \\
                        $m_\mathrm{gas}$ [u] & 2.02  & $m_\mathrm{TiO2}$ [u] & 79.87    \\
                                             &       &  $r_\mathrm{TiO2}$ [\r{A}] & 2.32 \\
                        \hline\hline
                \end{tabular}
    \end{table}

    \subsection{Internal temperature}
    
    \subsubsection{Collisional}
    \label{sec:RelaxTime_internal}
    
    Internal cooling or heating of clusters via elastic collisions with gas depends on the size ratio between cluster species and gas-phase molecules which is typically expressed with the Knudsen number \citep{woitke_dust_2003}:
    \begin{align}
        Kn = \frac{\Bar{l}}{2r_\mathrm{N}} ,
    \end{align}
    where $\Bar{l} = (\sigma_\mathrm{N} n_\mathrm{gas})^{-1}$ [cm] is the mean free path of the gas and $2r_\mathrm{N}$ is the diameter of the cluster (see also Table \ref{tab:tio2_data}). \\
    
    \noindent \underline{Case 1: $Kn \gg 1$}
    
    In the high Knudsen number limit, the theory of \citet{burke_gas-grain_1983} is used. Here, the cooling rate per unit volume $\Lambda_\mathrm{N}$ [erg s$^{-1}$cm$^{-3}$] of a cluster can be used to calculate the energy change for each particle per time (in the approximation $m_\mathrm{gas} \ll m_\mathrm{N}$):
    \begin{align}
        \nonumber \frac{dE^\mathrm{int}}{dt} &= \frac{\Lambda_\mathrm{N}}{n_\mathrm{N}} \approx \sqrt{\frac{8 k^3 }{\pi m_\mathrm{gas}}} \Bar{\alpha}_T \pi r_\mathrm{N}^2  n_\mathrm{gas} \sqrt{T_\mathrm{gas}} ( T_\mathrm{gas} - T_{N}^\mathrm{int}) ,
    \end{align}
    where $\Bar{\alpha}_T$ is the average accommodation coefficient. The exact value of $\Bar{\alpha}_T$ depends on the colliding species, the gas composition, and the temperature, but typically it is within 0.1 to 0.9 \citep{burke_gas-grain_1983}. To assess the timescales in orders of magnitude, we approximate the average accommodation coefficient as $\Bar{\alpha}_T = 0.5$.
    
    To calculate the collisional internal cooling timescale $\tau_\mathrm{gc}^\mathrm{int}$ towards the gas temperature, we need to relate the energy change of the cooling rate with the internal temperature change of the clusters. Using Eq. \ref{eq:Thermodynmaic_Temperature}, we find the following:
    \begin{align}
        \frac{dT_\mathrm{N}^\mathrm{int}}{dt} &= \frac{dT_\mathrm{N}^\mathrm{int}}{dE^\mathrm{int}} \frac{dE^\mathrm{int}}{dt} = (\tau_\mathrm{gc}^\mathrm{int})^{-1} (T_\mathrm{gas} - T_\mathrm{N}^\mathrm{int}) ,
    \end{align}
    where $\tau_\mathrm{gc}^\mathrm{int}$ is given by
    \begin{align}
        \label{eq:cooling_internal_collisional}
        (\tau_\mathrm{gc}^\mathrm{int})^{-1} &\approx \frac{2 \Bar{\alpha}_T}{D_f} n_\mathrm{gas} r_\mathrm{N}^2 \sqrt{\frac{8 \pi k T_\mathrm{gas}}{m_\mathrm{gas}}} .
    \end{align}
    Assuming a constant gas temperature $T_\mathrm{gas} (t) = T_\mathrm{gas}$, this system can be solved and leads to the following temperature evolution for a cluster after collision:
    \begin{align}
        T_\mathrm{N}^\mathrm{int} (t) = T_\mathrm{gas} + (T_\mathrm{N,0}^\mathrm{int} - T_\mathrm{gas}) \exp( - t/\tau_\mathrm{gc}^\mathrm{int}) ,
    \end{align}
    where $T_\mathrm{N,0}^\mathrm{int}$ [K] is the initial internal cluster temperature. Looking at our reference case of (TiO$_2$)$_2$ clusters in a gas as described in Table \ref{table:Standard_example}, we find a collisional cooling timescale for the (TiO$_2$)$_2$ internal temperature of $\tau^\mathrm{int}_\mathrm{gc} = 0.0059$ s. The kinetic cooling timescale is within an order of magnitude of the internal cooling timescale. This is not surprising, as Eq. \ref{eq:cooling_kinetic_collisional} and Eq. \ref{eq:cooling_internal_collisional} have the same $n_\mathrm{gas}$, $r_N,$ and $\langle \nu_r \rangle$ dependencies and only differ in their prefactor. This relation stems from the nature of collisional energy transfer. Kinetic energy is exchanged via elastic collisions of particles. The more often collisions occur and the more energy can be exchanged within one collision, the faster the kinetic temperature adjusts to the equilibrium temperature. Similarly, internal energy exchange can be described via collisions of gas particles with an internal 'spring' (vibration mode) of the cluster (see \citet{burke_gas-grain_1983}). Here, the timescale for thermal adjustment also depends on the collision rate and the energy amount exchange per collision. \\
    
    \noindent \underline{Case 2: $Kn \ll 1$}
    
    For larger clusters, the Knudsen number becomes increasingly small. In this case, we use the definition of the isobaric-specific heat and of the heat flow \citep[Eq. 15.19 and 17.1 of ][]{tipler_physik_2015}:
    \begin{align}
        \label{eq:def_cp}
        dQ &= m_\mathrm{N} c_p dT_\mathrm{N}^\mathrm{int} , \\
        \label{eq:def_heatFlow}
        \frac{dQ}{dt} &= \frac{k_c A_\mathrm{N}}{\Delta x} (T_\mathrm{gas} - T_\mathrm{N}^\mathrm{int}),
    \end{align}
    where $c_p$ [erg g$^{-1}$K$^{-1}$] is the specific heat capacity for constant pressure, $k_c$ [W cm$^{-1}$K$^{-1}$] is the thermal conductivity, and $A_\mathrm{N}$ [cm$^2$] is the cluster surface area. We assume that the energy exchange between the cluster and the gas happens over the mean free path of the gas ($\Delta x = \Bar{l}$). The thermal conductivity $k_c$ of an ideal gas is as follows \citep[Eq. 10-25 of ][]{sears_thermodynamics_1975}:
    \begin{align}
        \label{eq:thermal_conductivity}
        k_c = \frac{1}{3}~ c_V ~ \rho_\mathrm{gas} ~ \Bar{l} ~ \langle v \rangle ,
    \end{align}
    where $c_V$ [erg g$^{-1}$K$^{-1}$] is the isochoric-specific heat capacity. Using Eqs. \ref{eq:def_cp}, \ref{eq:def_heatFlow}, and \ref{eq:thermal_conductivity} leads to the following solution for the internal temperature:
    \begin{align}
        \frac{dT^\mathrm{int}}{dt} &= \frac{k_c ~A}{\Bar{l} ~m_\mathrm{N} ~c_p} (T_\mathrm{gas} - T_\mathrm{N}^\mathrm{int}) \\
        &\approx \frac{c_V ~ \rho_\mathrm{gas} ~ \pi r_\mathrm{N}^2}{3 c_p ~ m_\mathrm{N}} \sqrt{\frac{8 k T_\mathrm{gas}}{\pi m_\mathrm{gas}}} (T_\mathrm{gas} - T_\mathrm{N}^\mathrm{int}) , \\
        \label{eq:TempRelax}
        T_\mathrm{N}^\mathrm{int}(t) &= T_\mathrm{gas} + (T_\mathrm{N}^\mathrm{int}(0) - T_\mathrm{gas}) \exp(-t / \tau_\mathrm{gc}^\mathrm{int}) , \\
        (\tau_\mathrm{gc}^\mathrm{int})^{-1} &\approx \frac{c_V ~ m_\mathrm{gas}}{3 c_p ~ m_\mathrm{N}} n_\mathrm{gas} r_\mathrm{N}^2 \sqrt{\frac{8 \pi k T_\mathrm{gas}}{m_\mathrm{gas}}} .
    \end{align}

    \subsubsection{Radiative}
    \label{sec:RelaxTime_Radiative}
    
    Radiative cooling is especially important for a cluster with large dipole moments in regions where the gas density is low and collisional cooling processes become inefficient \citep{woitke_gas_1996, plane_master_2022}. Because of their intermediate size, small clusters are not well described as black bodies, but rather they radiate via discrete de-excitation (i.e. relaxation) of rotationally, vibrationally, and electronically excited states \citep{woitke_radiation_2009, coppola_radiative_2011, ferrari_radiative_2019}. Each cluster size $i$ has discrete energy levels $E_i^a$ [erg] with $a \geq 1$ and the ground state is denoted with G ($E_i^1 = E_i^G$). The energy levels have a number density of $n_i^a$ [cm$^{-3}$] which follow the Boltzmann distribution at the specific temperature $T^{int}_N$, 
    
    \begin{align}
        \frac{n_{i}^{a}}{n_{i}^{G}} &= \frac{g_{i,a}}{g_{i,G}}\exp \left( \frac{-(E_{i}^{a} - E_{i}^{G})}{k T_i^\mathrm{int}} \right) ,\\
        n_{i}^{G} &= n_i \left( \sum_{a=1}^{\infty} \frac{g_{i,a}}{g_{i,G}} \exp \left( \frac{-(E_{i}^{a} - E_{i}^{G})}{k T_N^\mathrm{int}} \right) \right)^{-1} ,
    \end{align}
    where $n_i = \sum_{a=1}^\infty n_i^a$ is the total number density of the $i$-mer and $g_{i,a}$ is the degeneracy of state $a$ of an $i$-mer. A relaxation from an upper level $a$ to a lower level $b$ of an $N$-mer changes the internal cluster energy $E_{i}^\mathrm{int}$ by
    \begin{align}
        \Eval{\frac{dE_{i}^\mathrm{int}}{dt}}{ab}{}= - A_{i}^{ab} (E_{i}^{a} - E_{i}^{b}) \frac{n_{i}^{a}}{n_i}  ,
    \end{align}
    where $A_{i,ab}$ [s$^{-1}$] is the Einstein coefficient. The internal temperature change is then given as the sum over all changes in energy levels:
    \begin{align}
        \nonumber \frac{dT_i^\mathrm{int}}{dt} &= \sum_{a=2}^{\infty} \sum_{b=1}^{a-1} \Eval{\frac{dT^\mathrm{int}_i}{dE_{i}^\mathrm{int}} \frac{dE_{i}^\mathrm{int}}{dt}}{ab}{} \\
        \nonumber &= \frac{-2}{k D_i^f} \sum_{a=2}^{\infty} \sum_{b=1}^{a-1} A_{i}^{ab} (E_{i}^{a} - E_{i}^{b}) \frac{n_{i}^{a}}{n_i}  \\
        \label{eq:GeneralRadiativeCooling}
        &= \frac{- 2}{k D_i^f}\frac{n_{i}^{G}}{n_i} \sum_{a=2}^{\infty} \sum_{b=1}^{a-1}  A_{i}^{ab} (E_{i}^{a} - E_{i}^{b}) \frac{g_{i,a}}{g_{i,G}} \exp \left( \frac{- (E_{i}^{a} - E_{i}^{G})}{k T_i^\mathrm{int}} \right) .
    \end{align}
    
    The investigation of spontaneous and stimulated radiative emissions of clusters, that is to say their Einstein coefficients, is challenging. Previous studies used strong assumptions to approximate the thermal non-equilibrium caused by radiative processes \citep{nuth_vibrational_1981, nuth_joseph_a_iii_silicates_2006}. Recently, a more detailed investigation was done by \citet{plane_master_2022}. They calculated the Einstein coefficients of the potentially dust-forming silicate OSi(OH)$_2$ at the B3LYP/6-311+g(2d,p) level of theory assuming that harmonic vibrations and Einstein coefficients remain constant in each of the vibration modes. They found that for gas densities below $10^{12}$ cm$^{-3}$ at $T_\mathrm{gas} = 1583$ K, the vibrational (internal) non-equilibrium can reach as large as $T_\mathrm{gas} - T^\mathrm{vib}_{\mathrm{OSi(OH)_2}} \approx 900$ K. The corresponding dissociation rate is reduced by several orders of magnitude. We note, however, that clusters can show significant anharmonic vibrations and temperature dependencies which further complicate the exact descriptions of radiative emissions \citep{guiu_how_2021}. Owing to these complications we do not explicitly derive master equations in internal non-equilibirum, but account for its effect by adopting different internal cluster temperatures.

\section{Kinetic nucleation in thermal non-equilibrium}
\label{sec:Testing}
    
    To test our equations in thermal equilibrium, a comparison to the work of \citet{lee_dust_2015} and \citet{boulangier_developing_2019} is presented in Section \ref{sec:Testing_Teq}. In Section \ref{sec:Testing_Tdis_effects}, different temperature non-equilibria and their impact on kinetic cluster nucleation are analysed. In Section \ref{sec:Testing_eqvsneq}, we compare the effect of thermal non-equilibrium with thermal equilibrium on the cluster number densities.

    \subsection{Kinetic nucleation in thermal equilibrium}
    \label{sec:Testing_Teq}
    
\begin{figure}
      \centering
      \includegraphics[width=\hsize]{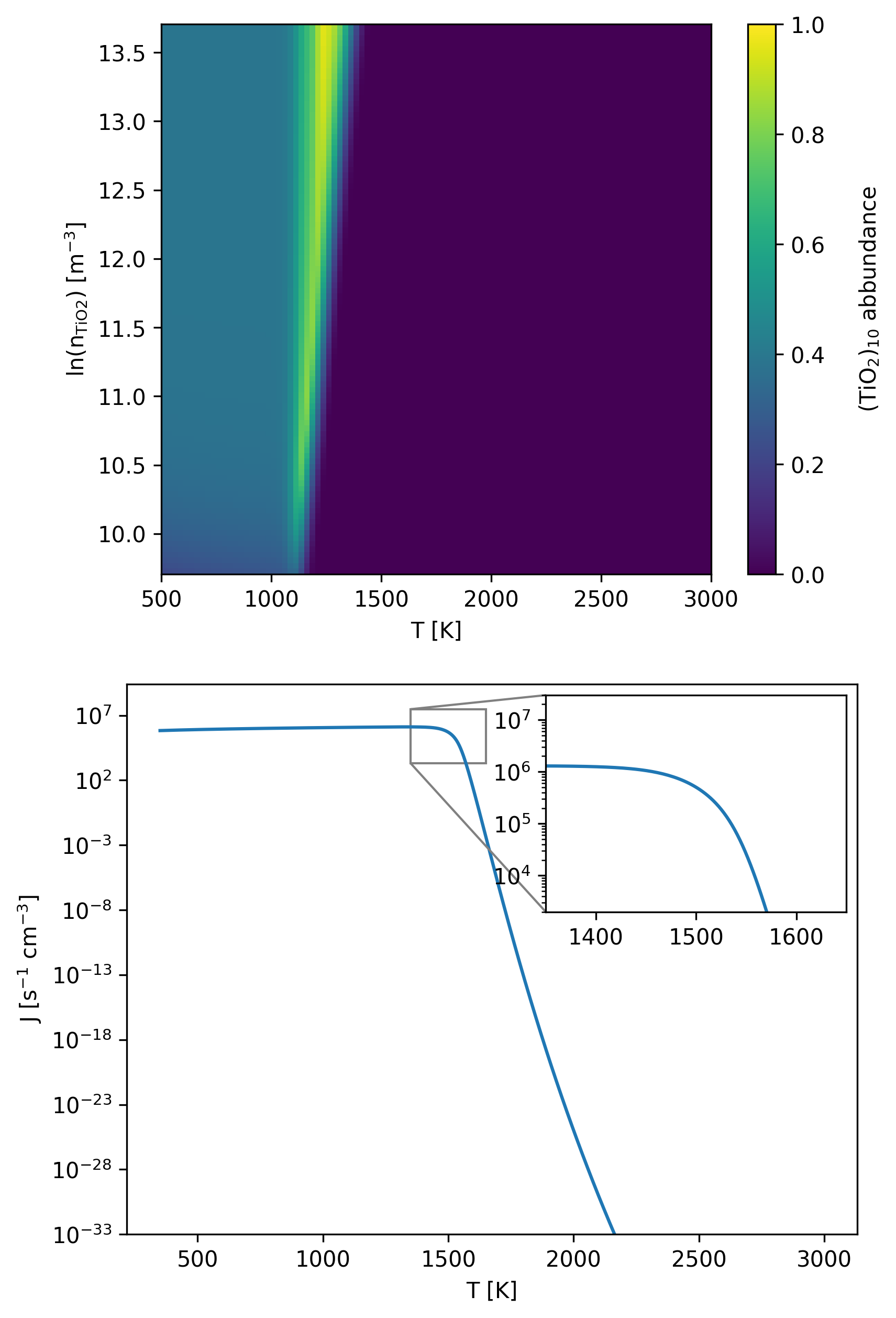}
      \caption{Comparison to previous studies. \textbf{Top:} (TiO$_2$)$_{10}$ abundance after 1 year with respect to the total number density of all clusters $\sum_{i=1}^{10} n_\mathrm{(TiO_2)_i}$. This graph was made to be compared with the results of \citet{boulangier_developing_2019}. \textbf{Bottom:} Nucleation rate of TiO$_2$. The simulation was conducted for an initial uniform cluster number density of $n_\mathrm{(TiO_2)_N} = 10^{9}$ cm$^{-3}$ and evaluated after $t = 10^5$ s. This graph was made to be compared with the results of \citet{lee_dust_2015}.}
      \label{fig:Test_equi}
  \end{figure}

    First we compare our model to that of \citet{boulangier_developing_2019}. Similar to their analysis, we use (TiO$_2$)$_N$ clusters up to a size of $N=10$ and evolve the chemical network for a period of $t = 1$ yr. Instead of number densities, they used mass densities of the total gas, which is related to the gas number densities by the mean molecular weight. Furthermore, they assumed in their closed nucleation model that all Ti is bound in TiO$_2$. To reproduce their results, we calculated the initial TiO$_2$ number density using the molecular weight of Ti ($m_\mathrm{Ti} = 47,867$ u $\approx 8 \times 10^{-23}$ g) and the Ti mass fraction ($\chi_\mathrm{Ti} = 2.84 \times 10^{-6}$). We assume thermal equilibrium between the gas and all clusters ($T = T_{gas} = T_{i}^\mathrm{int} = T_{i}^\mathrm{kin}$ for all cluster sizes $i$). Using this conversion\footnote{The conversion is given by $n_\mathrm{TiO_2} = \rho_\mathrm{gas} ~\chi_\mathrm{Ti} ~/~ m_\mathrm{Ti} \approx 3.6 \times 10^{20} ~\mathrm{g^{-1}} ~\rho_\mathrm{gas}$.}, we calculated the relative (TiO$_2$)$_{10}$ abundance, defined as
    \begin{align}
        \xi_\mathrm{(TiO_2)_{10}} = \frac{n_\mathrm{(TiO_2)_{10}}}{\sum_{i=1}^{10} n_\mathrm{(TiO_2)_{i}}},
    \end{align}
    after 10$^5$ seconds for a temperature range of 500 K to 3000 K and a number density range of roughly 5.08$\times$10$^3$ cm$^{-3}$ to 5.08$\times$10$^{7}$ cm$^{-3}$ (values taken from \citet{boulangier_developing_2019}). Our result can be seen in the top panel of Fig. \ref{fig:Test_equi}. Overall, our (TiO$_2$)$_{10}$ abundances match the results from \citet{boulangier_developing_2019} which was expected because both use the same underlying assumptions and the same setup for the nucleation network. 
    
    Next, we compare our network to the nucleation model of \citet{lee_dust_2015}. In their study, the nucleation rate is calculated for homo-molecular monomer nucleation. They consider clusters of sizes up to (TiO$_2$)$_{10}$. Since our work also uses polymer nucleation, we calculate the nucleation rate by instantly dissociating\footnote{This process is also known as a Maxwell-Demon.} all (TiO$_2$)$_{10}$ clusters into ten monomers (10$\times$TiO$_2$). This leads to a constant nucleation flux. Our results are shown in the bottom of Fig. \ref{fig:Test_equi}. Above 600 K, our nucleation network (Eq. \ref{eq:chemical_network}) produces comparable nucleation rates as the non-classical nucleation rate of \citet{lee_dust_2015}. Below 600 K, they found a steep decrease in the nucleation rate which is not reproduced by our model showing a constant rate of roughly $J = 1 \times 10^{6}$ s$^{-1}$ cm$^{-1}$. Comparing the results of \citet{lee_dust_2015} with \cite{boulangier_developing_2019} shows that both predict a lack of (TiO$_2$)$_{10}$ for homo-molecular monomer nucleation. Polymer nucleation, on the other hand (as implemented in our network and the one from \cite{boulangier_developing_2019}), showed non-negligible (TiO$_2$)$_{10}$ abundances below 1000 K.

   \subsection{Effect of thermal non-equilibrium}
   \label{sec:Testing_Tdis_effects}
   
   \begin{figure*}
       \centering
       \includegraphics[width=\hsize]{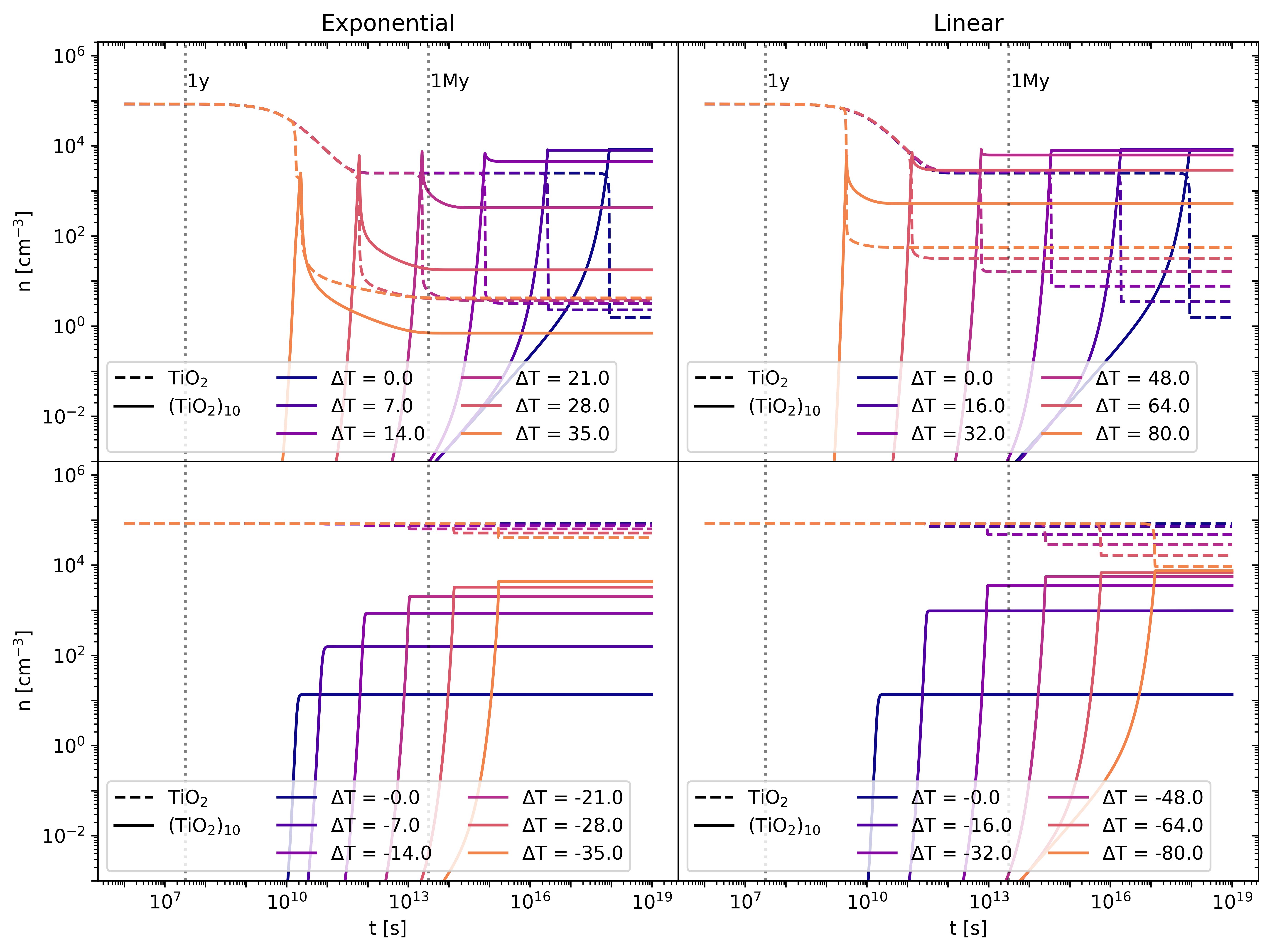}
       \caption{(TiO$_2$)$_N$ cluster number densities as a function of time under kinetic-to-gas thermal offsets ($T_N^\mathrm{kin} = T_N^\mathrm{int} \neq T_\mathrm{gas}$). The initial TiO$_2$ number density is $n_\mathrm{TiO_2} = 10^{4}$ cm$^{-3}$ in a H$_2$ gas in thermal equilibrium. \textbf{Top left:} Exponential temperature offsets (see Eq. \ref{eq:exponential_offsets}) at $T_\mathrm{gas} = 1000$ K. \textbf{Bottom left:} Exponential temperature offsets at $T_\mathrm{gas} = 1250$ K. \textbf{Top right:} Linear temperature offsets (see Eq. \ref{eq:linear_offsets}) at $T_\mathrm{gas} = 1000$ K. \textbf{Bottom right:} Linear temperature offsets at $T_\mathrm{gas} = 1250$ K.}
       \label{fig:Test_Tdis}
   \end{figure*}
   
   Thermal non-equilibrium affects the kinetic nucleation network in two ways: firstly, via the relative velocity distribution $f_r(\nu_r)$. If the velocity distribution of the colliding cluster is a Maxwell-Boltzmann distribution, the temperature dependence is given within the TCM-reduced mass $\mu_T$. In the general case, the velocity distribution can depend on the type of non-equilibrium present. For this section, we assume the velocity distribution of all clusters to be a Maxwell-Boltzmann distribution. Secondly, the backward rates (Eq. \ref{eq:Gibbs_eq_general}) depend on the kinetic and internal temperatures of the clusters. The dependencies can be divided as follows:
   \begin{align}
       k^- &= k^+ \frac{p^\ominus}{k T_\mathrm{gas}} A ~B ~C , \\
       A &= \exp \left(  \sum_{i \in \zeta} \frac{\delta(i)}{k T_i^\mathrm{kin}} \Big[ \mu_i^\ominus (T_\mathrm{gas}) - i \mu_1^\ominus(T_i^\mathrm{kin}) + k (T_i^\mathrm{kin} - T_\mathrm{gas})  \Big] \right) , \\
       B &= \exp \left(\sum_{i \in \zeta} \frac{\delta(i)}{k T_i^\mathrm{kin}} \omega_i (T_i^\mathrm{kin}, T_i^\mathrm{int}) \right) , \\
       C &= \left( \frac{k T_\mathrm{gas} \mathring{n}_\mathrm{TiO_2}}{p^\ominus} \right)^{- \sum_{i \in \zeta} \delta(i) i \frac{T_\mathrm{gas}}{ T_i^\mathrm{kin}}} ,
   \end{align}
   where $A$ is the correction term due to kinetic-to-gas thermal non-equilibrium, $B$ the correction term due to kinetic-to-internal thermal non-equilibrium, and $C$ the correction due to the Lagrangian multiplier. 
   
   The correction term $C$ depends on the equilibrium number density of the monomer $\mathring{n}_\mathrm{TiO_2}$, which in turn depends on the reaction rates. To decouple this dependency, we start by assuming $C = 1$. Afterwards, we calculate the maximum offset to check if this assumption holds. The correction term $A$ only depends on the difference between kinetic temperature and gas temperature. If the clusters are in kinetic-to-gas thermal equilibrium ($T^\mathrm{kin}_N = T_\mathrm{gas}$), $A$ becomes the exponent of the backward reaction rate for thermal equilibrium (as in e.g. \citet{boulangier_developing_2019}). The correction term $B$ does not depend on the temperature of the gas phase, but rather on the thermal difference between internal and kinetic temperature. As long as N-mers are in kinetic-to-internal thermal equilibrium ($T^\mathrm{kin}_N = T^\mathrm{int}_N$), the correction term $B$ is equal to 1.
   
   In this section we assume kinetic-to-internal thermal equilibrium for all clusters to study thermal non-equilibrium between gas phase and dust. This assumption is similar to the thermal non-equilibrium considered in the model of \citet{patzer_dust_1998}, \citet{helling_dust_2006}, and \citet{kohn_dust_2021}. Our TiO$_2$ reference case therefore allows for the importance of thermal non-equilibrium for their models to be estimated. Studying the effect kinetic-to-internal non-equilibrium, similar to \citet{plane_master_2022}, would also be interesting. Unfortunately, this requires evaluating the Gibbs free energy of clusters in kinetic-to-internal non-equilibrium which is outside of the scope of this paper and will be dealt with in a separate study.
   
   To investigate the effect of the correction terms $A$, we use four different simulations. We assume kinetic-to-internal thermal equilibrium ($T_N^\mathrm{int} = T_N^\mathrm{kin} \neq T_\mathrm{gas}$) and use two temperature structures each: 
   \begin{align}
        \label{eq:exponential_offsets}
       &\quad \text{Exponential: } &T_{N}^\mathrm{kin} &= T_\mathrm{gas} + \frac{e^{N-1} - 1}{e^9 - 1} \Delta T , \\
        \label{eq:linear_offsets}
       &\quad \text{Linear: } &T_N^\mathrm{kin} &= T_\mathrm{gas} + \frac{(N-1)}{9} \Delta T ,
   \end{align}
   where $\Delta T$ is a free parameter quantifying the kinetic thermal non-equilibrium between the monomer (n=1) and the decamer (n=10). The definitions were chosen so that $\Delta T = T^\mathrm{kin}_\mathrm{(TiO_2)_{10}} - T^\mathrm{kin}_\mathrm{TiO_2}$ holds in all cases. Both offsets are toy models to show the effect of different offsets on the resulting number densities.
   
   The simulations use TiO$_2$ as nucleating species with an initial monomer density $n_\mathrm{TiO_2} = 10^{4}$ cm$^{-3}$ in a H$_2$ gas with a density of $n_{gas} = 10^{12}$ cm$^{-3}$ at $T_\mathrm{gas} = 1000$ K. Clusters of size $N \leq 4$ are considered to associate and disassociate via an additional collision partner M. We used temperature offsets in the range of $0~\mathrm{K} \leq \Delta T \leq 35 ~\mathrm{K}$ for the exponential offset (top-left panel of Fig. \ref{fig:Test_Tdis}) and $0~\mathrm{K} \leq \Delta T \leq 80~\mathrm{K}$ for the linear offset (top-right panel of Fig. \ref{fig:Test_Tdis}). The simulations show a decrease in (TiO$_2$)$_{10}$ number density with increased temperature offsets. This decrease is stronger for the exponential offset than for the linear offset. Therefore, not only does the general temperature increase, but the type of non-equilibrium also affects the resulting number densities. For $T_{gas}$=1000 K, (TiO$_2$)$_{10}$ does not represent the most abundant cluster in all cases. As such, (TiO$_2$)$_8$ becomes the most abundant cluster size for exponential offsets of $\Delta T = 21$ K and for linear offsets of $\Delta T = 80$ K. The peaks and declines in (TiO$_2$)$_{10}$ number density for increased thermal offsets are caused by the initially efficient growth reactions due to the high abundance of small cluster and the enhanced dissociation rate, respectively. A similar overshoot can be seen in \citet{kohn_dust_2021} who analysed different evaporation efficiencies for TiO$_2$ nucleation.
   
   We also considered the reverse temperature offset for both simulations. For this, we repeated the simulation with $T_\mathrm{gas} = 1250$ K where, in thermal equilibrium, the (TiO$_2$)$_{10}$ number density is significantly lower than at $T_\mathrm{gas} = 1000$ K. The results for exponential offsets can be seen in the bottom-left panel of Fig. \ref{fig:Test_Tdis}. The results for linear offsets can be seen in the bottom-right panel of Fig. \ref{fig:Test_Tdis}. In both simulations, the (TiO$_2$)$_{10}$ number density increases with decreased kinetic temperatures. Overall, we find that for TiO$_2$ around 1000 to 1250 K lower kinetic temperatures favour nucleation and higher kinetic temperatures hamper nucleation.
   
   To evaluate the assumption that $C \approx 1$, we compare $C$ to the uncertainty factors $F_k$ of the reaction rates \citep{baulch_evaluated_1992, dobrijevic_effect_1998}. The upper and lower limit for $C$ throughout all simulations are $0.962 < C \leq 1.0,$ which corresponds to a correction factor of $F_C = 1.04$. Typical uncertainty factors of hydrocarbon reactions at $T_\mathrm{gas} < 300$ K exceed 1.5. \citep{dobrijevic_effect_1998, dobrijevic_effect_2003, hebrard_photochemical_2006}. \citet{hebrard_determining_2015} extrapolate the uncertainty factor to temperatures up to 1000 K and estimate that the uncertainty factors exceed 1.26 for bi-molecular reactions. Therefore, the assumption of $C \approx 1$ is justified.
   
   In all simulations, thermal non-equilibrium can change the (TiO$_2$)$_{10}$ number density by several orders of magnitude. The exact change in the abundance of the largest cluster depends on the gas temperature $T_\mathrm{gas}$, the clustering species, the cluster number densities $n_N$, and the type of thermal non-equilibrium present.

   \subsection{Thermal equilibrium versus non-equilibrium}
   \label{sec:Testing_eqvsneq}
   
   In Section \ref{sec:Testing_Tdis_effects}, we have shown that positive cluster temperature offsets can lead to lower (TiO$_2$)$_{10}$ number densities and negative cluster temperature offsets can lead to higher (TiO$_2$)$_{10}$ number densities. Since this behaviour is also expected for thermal equilibrium, we compare the change in (TiO$_2$)$_{10}$ number density for thermal equilibrium to thermal non-equilibrium. We choose an H$_2$ gas with a density of $n_{gas} = 10^{12}$ cm$^{-3}$, an initial TiO$_2$ monomer density $n_\mathrm{TiO_2} = 10^{4}$ cm$^{-3}$, and assume exponential temperature offsets (see Eq. \ref{eq:exponential_offsets}).
   
\begin{figure}
      \centering
      \includegraphics[width=\hsize]{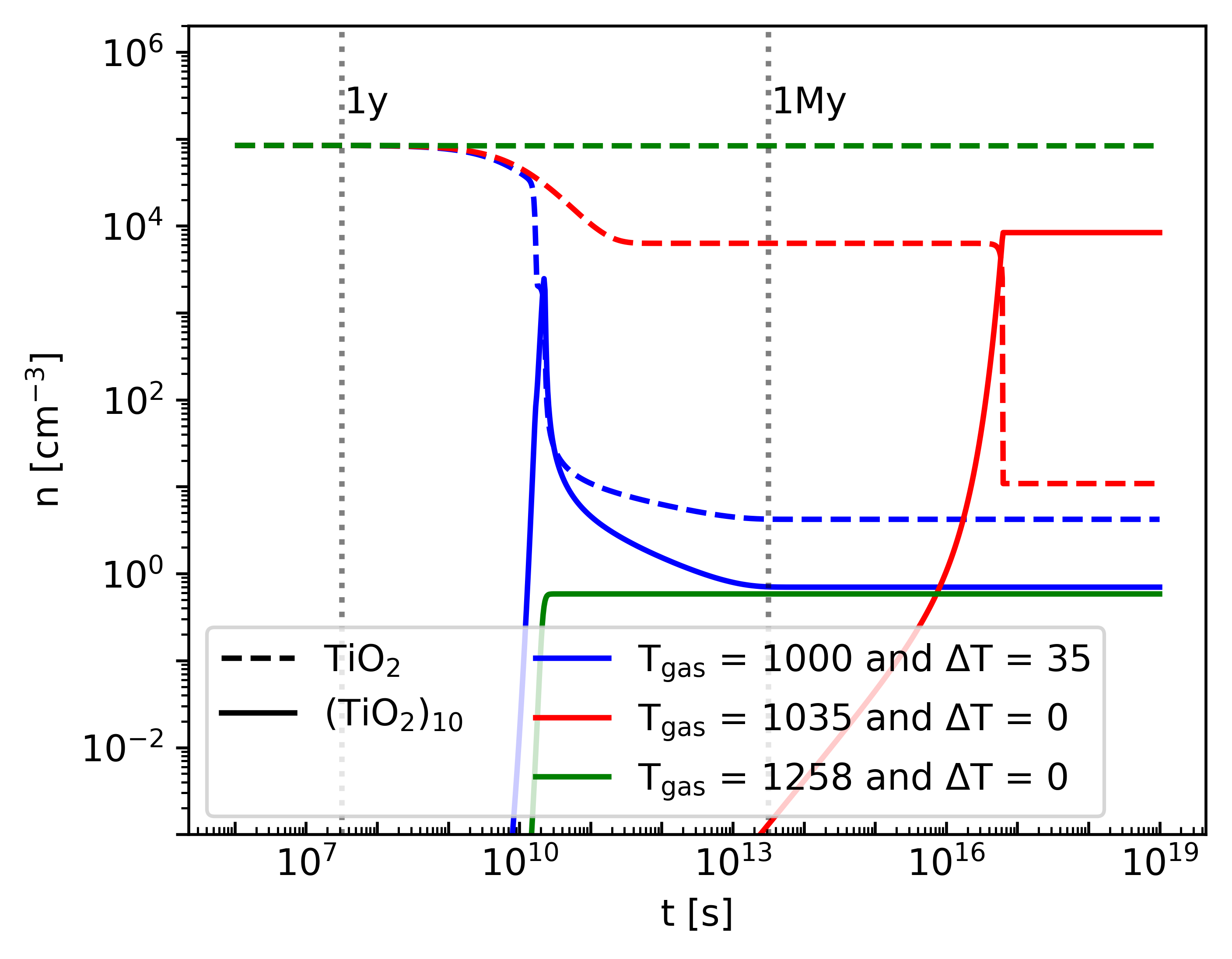}
      \caption{Comparison of TiO$_2$ and (TiO$_2$)$_{10}$ number densities under thermal equilibrium ($\Delta T = 0$ K) and kinetic-to-gas thermal non-equilibrium ($\Delta T \neq 0$ K) conditions.}
      \label{fig:Test_eq_to_neq}
  \end{figure}

   The first simulation is done in thermal non-equilibrium using a gas temperature of $T_\mathrm{gas} = 1000$ K and a temperature offset\footnote{$\Delta T = 35$ K was chosen because it showed a clear impact on the (TiO$_2$)$_{10}$ number density.} of $\Delta T = 35$ K for the clusters (see Section \ref{sec:Testing_Tdis_effects}). The second simulation is done assuming thermal equilibrium ($\Delta T = 0$ K) at $T_\mathrm{gas} = T_N^\mathrm{kin} = T_N^\mathrm{int} = 1035$ K. The results can be seen in Fig. \ref{fig:Test_eq_to_neq}.
   
   Even though the (TiO$_2$)$_{10}$ clusters are at the same temperature, the (TiO$_2$)$_{10}$ number densities are 4 orders of magnitude smaller if thermal non-equilibrium is present. We tested additional gas temperatures assuming thermal equilibrium and found that to reach a similar (TiO$_2$)$_{10}$ number density as in the thermal non-equilibrium case, the gas temperature has to be around $T_\mathrm{gas} = T_N^\mathrm{kin} = T_N^\mathrm{int} = 1258$ K.

\section{Conclusion}
\label{sec:Conclusion}

    We developed a kinetic nucleation model describing cluster formation and destruction in thermal non-equilibrium using non-identical temperatures for the clusters and the gas. This model does not rely on assumptions made by previous studies of thermal non-equilibrium nucleation frameworks. Our model includes realistic termolecular association and collisional dissociations for the smallest cluster sizes. To derive a dissociation rate in thermal non-equilibrium, we make use of the law of mass action, the principle of detailed balance, and the minimization of the Gibbs free energy in a general form accounting for non-equilibrium effects. All particles, which are not part of the nucleation process, are assumed to be in thermal equilibrium at the temperature $T_\mathrm{gas}$. Clusters are described by a kinetic (transitional) temperature $T_N^\mathrm{kin}$ and an internal temperature $T_N^\mathrm{int}$. All temperatures can differ from each other.
    
    Thermal non-equilibrium can affect the synthesis of larger TiO$_2$ clusters. Lower cluster temperatures lead to an increased abundance of larger clusters. This relation was already predicted by previous studies and our simulations confirm and quantify the impact of thermal non-equilibrium on cluster formation. For TiO$_2$ at $n_\mathrm{TiO_2} = 10^{4}$ cm$^{-3}$ in a H$_2$ gas at $n_\mathrm{H_2} = 10^{12}$ cm$^{-3}$ and $T_\mathrm{gas} = 1000$ K, we found that the number density of (TiO$_2$)$_{10}$ can decrease over an order of magnitude for kinetic-to-gas temperature offsets with $(T^\mathrm{kin}_\mathrm{(TiO_2)_{10}} - T_\mathrm{gas}) \leq 21$ K. For a higher gas temperature of $T_\mathrm{gas} = 1250$ K, we found over an order of magnitude increase in the (TiO$_2$)$_{10}$ number density for kinetic-to-gas temperature offsets of $(T^\mathrm{kin}_\mathrm{(TiO_2)_{10}} - T_\mathrm{gas}) \leq -14$ K.
    
    In environments where thermal non-equilibrium is already observed or suspected, such as the outflows of AGB stars, it is crucial to consider the effect of thermal non-equilibrium on kinetic nucleation. Already small temperature offsets within clustering species can cause significant changes in the abundance of larger clusters. The model derived in this work can be used to add thermal non-equilibrium considerations to chemical (kinetic) networks.

\begin{acknowledgements}
    The authors thank Julian Lang for his contribution and help to this work. S.K., L.D and C.H. acknowledges funding from the European Union H2020-MSCA-ITN-2019 under grant agreement no. 860470 (CHAMELEON). L.D and D.G. acknowledges support from the ERC consolidator grant 646758 AEROSOL. D.G. acknowledges support from the project grant ``The Origin and Fate of Dust in the Universe'' from the Knut and Alice Wallenberg foundation.
\end{acknowledgements}

\bibliographystyle{aa} 
\bibliography{bibliography} 

\begin{appendix}

    \section{Cluster data}
    
    \begin{table*}
        \centering
        \caption{Cluster data for (TiO$_2$)$_N$ clusters, including: geometrical (Geo) radius, Van-der-Waals (VdW) radius, and mass.}
        \label{tab:tio2_data}
        \begin{tabular}{l l l l l l l l l l l }
            \hline\hline
             cluster properties & N=1 & N=2 & N=3 & N=4 & N=5 & N=6 & N=7 & N=8 & N=9 & N=10 \\
             \hline
             radius (Geo) [\AA] & 0 & 0.90 & 1.40 & 1.77 & 2.04 & 2.27 & 2.54 & 2.77 & 2.95 & 3.18 \\
             radius (VdW) [\AA] & 2.32 & 2.81 & 3.14 & 3.41 & 3.66 & 3.87 & 4.06 & 4.27 & 4.39 & 4.57 \\
             mass [u] & 79.9 & 159.7 & 239.6 & 319.4 & 399.3 & 479.2 & 559.1 & 638.9 & 718.8 & 798.7 \\
             \hline
        \end{tabular}
    \end{table*}
    
    \label{sec:Appendix_Gibbs_TiO2_data}
    In this section we present the TiO$_2$ data used throughout the paper. \\

    \noindent \underline{Cluster radius}
    
\begin{figure}
       \centering
       \includegraphics[width=\hsize]{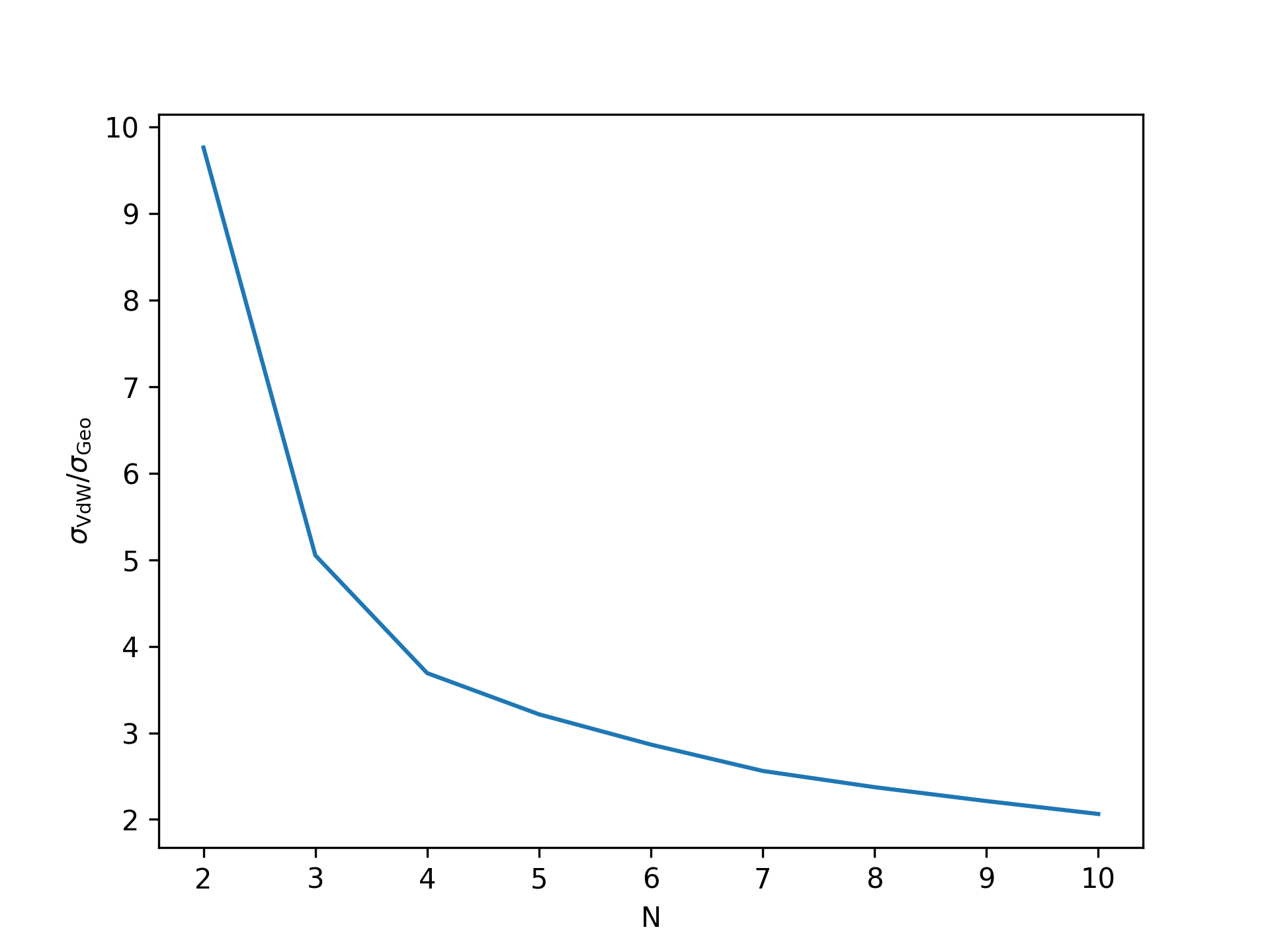}
       \caption{Ratio of the Van der Waals (VdW) and the geometrical (Geo) cross section for different cluster sizes.}
       \label{fig:RadiusCompare}
   \end{figure}

    The radius of small TiO$_2$ clusters cannot be calculated in a straightforward manner, owing to the diverse and non-identical cluster shapes (i.e. geometries). However, a cluster volume can be derived by using either the coordinates of the atomic cores (i.e. purely geometrical), or by including Van der Waals volumes accounting for the presence of electrons. Assuming sphericity, a cluster radius can be deduced. Both sets of values are shown in Table \ref{tab:tio2_data}. The difference between these two on the resulting cross section can be seen in Fig. \ref{fig:RadiusCompare}. For the calculations in Section \ref{sec:RelaxTime} and \ref{sec:Testing}, we used the radius calculated using Van der Waals forces. Readers can also refer to \citet{kohn_dust_2021} for an additional discussion on cluster radii. \\
    
    \noindent \underline{Gibbs free energy}
    
    The Gibbs free energies of formation of the TiO$_2$ clusters were taken from \cite{sindel_revisiting_2022}. \\

    \noindent \underline{Reaction energy}
    
    The reaction energy E$_{reac}$ of the TiO$_2$ dimerisation (i.e. TiO$_2$+TiO$_2$ $\rightarrow$ (TiO$_2$)$_2$) was calculated with density functional theory using the software package Gaussian16 \citep{frisch_gaussian16_2016}. The calculations were performed at the B3LYP/cc-pVTZ level of theory \citep{becke_densityfunctional_1993} including the vibrational zero-point correction:
    \begin{align}
     \nonumber E_{reac} &= E(\mathrm{(TiO_2)_2}) - 2 \times E (\mathrm{TiO_2})  \\
     &\simeq 7.86 10^{-19} J.
    \end{align}

    \noindent \underline{Dissociation reactions}
    
    All collisional dissociation reactions for TiO$_2$ considered in Section \ref{sec:Testing_Tdis_effects} are listed in Table \ref{table:3_body_tio2}. The termoelcular association rates are calculated using the dissociation rate and detailed balance as described in Section \ref{sec:Model_3body}. The thermal non-equilibrium correction factor $q_i$ is defined as follows:
    \begin{align}
        q_i = \sqrt{\frac{m_\mathrm{gas} T^\mathrm{kin}_i + m_i T_\mathrm{gas}} {(m_\mathrm{gas} + m_i) T_\mathrm{gas}}}.
    \end{align}
    Example values for the reaction rate coefficients of three-body association reactions of TiO$_2$ are shown in Table \ref{table:3_body_association_tio2}.
   
    \begin{table*}
        \caption{Collisional dissociation reactions of TiO$_2$ for cluster sizes $N \leq 4$ including thermal non-equilibrium effects.}
        \label{table:3_body_tio2}      
        \centering 
                \begin{tabular}{l l l}
                        \hline\hline
                        Reaction & Reaction rate coefficient [cm$^{3}$ s$^{-1}$]& References\\
                        \hline
                        (TiO$_2$)$_2$ + M $\rightarrow$ TiO$_2$ + TiO$_2$ + M & $1.4 \times 10^{-4} \exp (-48870/T^\mathrm{int}_\mathrm{(TiO_2)_2}) ~q_\mathrm{(TiO_2)_2}$ & \citet{plane_nucleation_2013} \\
                        (TiO$_2$)$_3$ + M $\rightarrow$ (TiO$_2$)$_2$ + TiO$_2$ + M & $1.4 \times 10^{-9} \exp (-62411/T^\mathrm{int}_\mathrm{(TiO_2)_3}) ~q_\mathrm{(TiO_2)_2}$ & Estimate from CCSD(T) \\
                        (TiO$_2$)$_4$ + M $\rightarrow$ (TiO$_2$)$_3$ + TiO$_2$ + M & $1.4 \times 10^{-9} \exp (-53569/T^\mathrm{int}_\mathrm{(TiO_2)_4}) ~q_\mathrm{(TiO_2)_2}$ & Estimate from CCSD(T) \\
                        (TiO$_2$)$_4$ + M $\rightarrow$ (TiO$_2$)$_2$ + (TiO$_2$)$_2$ + M & $1.4 \times 10^{-9} \exp (-57194/T^\mathrm{int}_\mathrm{(TiO_2)_4}) ~q_\mathrm{(TiO_2)_2}$ & Estimate from CCSD(T) \\
                         \hline\hline
                \end{tabular}
    \end{table*}
   
    \begin{table*}
        \caption{Recombination rate coefficients for the three-body association reactions of TiO$_2$ clusters up to size $N \leq 4$ including thermal non-equilibrium effects. All rates are calculated in kinetic-to-internal thermal equilibrium using exponential temperature offsets (See Eq. \ref{eq:exponential_offsets}).}
        \label{table:3_body_association_tio2}      
        \centering 
                \begin{tabular}{l l l}
                        \hline\hline
                        Reaction & Reaction rate coefficient [cm$^{6}$ s$^{-1}$] & Reaction rate coefficient [cm$^{6}$ s$^{-1}$]\\
                                 & for $T_\mathrm{gas} = 1000$ K and $\Delta T = 35$ K & for $T_\mathrm{gas} = 1250$ K and $\Delta T = -35$ K \\
                        \hline
                        TiO$_2$ + TiO$_2$ + M $\rightarrow$ (TiO$_2$)$_2$ + M             & $5.580 \times 10^{-28}$ & $1.219 \times 10^{-28}$ \\
                        (TiO$_2$)$_2$ + TiO$_2$ + M $\rightarrow$ (TiO$_2$)$_3$ + M       & $1.107 \times 10^{-37}$ & $1.788 \times 10^{-37}$ \\
                        (TiO$_2$)$_3$ + TiO$_2$ + M $\rightarrow$  (TiO$_2$)$_4$ + M      & $1.074 \times 10^{-38}$ & $1.741 \times 10^{-38}$ \\
                        (TiO$_2$)$_2$ + (TiO$_2$)$_2$ + M $\rightarrow$ (TiO$_2$)$_4$ + M & $4.318 \times 10^{-39}$ & $7.117 \times 10^{-39}$ \\
                        \hline\hline
                \end{tabular}
    \end{table*}

\end{appendix}

\end{document}